\documentclass[twocolumn,showpacs,preprintnumbers,amsmath,amssymb,prl]{revtex4-1}

\usepackage{graphicx}
\usepackage{dcolumn}
\usepackage{bm}

\begin{document}

\title{
Measuring the temperature dependence of individual two-level systems by direct coherent control
}
\author{J.~Lisenfeld$^{1,4}$, C.~M\"uller$^{2,4}$, J.~H.~Cole$^{3,4}$, P.~Bushev$^{1,4}$, A. Lukashenko$^{1,4}$, A.~Shnirman$^{2,4}$, and A.~V.~Ustinov$^{1,4}$}

\email{alexey.ustinov@kit.edu}
\affiliation{$^1$Physikalisches Institut, Karlsruhe Institute of Technology, D-76128 Karlsruhe, Germany\\
$^2$  Institut f\"{u}r Theorie der Kondensierten Materie, Karlsruhe Institute of Technology, D-76128 Karlsruhe, Germany\\
$^3$ Institut f\"ur Theoretische Festk\"{o}rperphysik, Karlsruhe Institute of Technology, D-76128 Karlsruhe, Germany\\
$^4$ DFG-Center for Functional Nanostructures (CFN), D-76128 Karlsruhe, Germany}

\email{alexey.ustinov@kit.edu}
\date{\today}

\begin{abstract}
	We demonstrate a new method to directly manipulate the state of individual two-level systems (TLSs) in phase qubits. 
	It allows one to characterize the coherence properties of TLSs using standard microwave pulse sequences, while the qubit is used only for state readout. 
	We apply this method to measure the temperature dependence of TLS coherence for the first time. 
	The energy relaxation time $T_1$ is found to decrease quadratically with temperature for the two TLSs studied in this work, 
	while their dephasing time measured in Ramsey and spin-echo experiments is found to be $T_1$ limited at all temperatures.
\end{abstract}


\maketitle

In the early 1970s, measurements of the thermal properties of amorphous materials \cite{Pohl71} led to the development of a phenomenological model to explain their specific heat and thermal conductivity at low temperature. Anderson, Halperin and Varma \cite{AndersonHalperinVarma71}, 
as well as Phillips \cite{Phillips72}, suggested the presence of an ensemble of two-level systems in the amorphous material, originating 
from quantum tunneling of individual atoms or a small group of atoms between two metastable lattice positions. 

The tunneling model was intensively tested experimentally by ensemble measurements performed on samples having a large TLS density, such as glasses. As an example, the life time of thermally exited states could be measured from the heat release of a sample after a rapid cool-down. Also, quantum coherent measurements of the decoherence times of (near-) resonantly excited subsets of TLSs were performed by monitoring the response to acoustic \cite{GraebnerGolding79} or electric echo pulses. 
Interpretation of these experiments inherently requires a statistical analysis of an inhomogeneous ensemble of TLSs, which is characterized by a distribution of dipole orientations and strengths as well as a spread of local strain fields. Since the microscopic nature of TLSs remains an actively debated topic, it is crucial to gain a deeper understanding by observing the properties of \emph{individual} TLSs and hereby raise the veil imposed by averaging.

\begin{figure}[b]
	\includegraphics[width=\columnwidth]{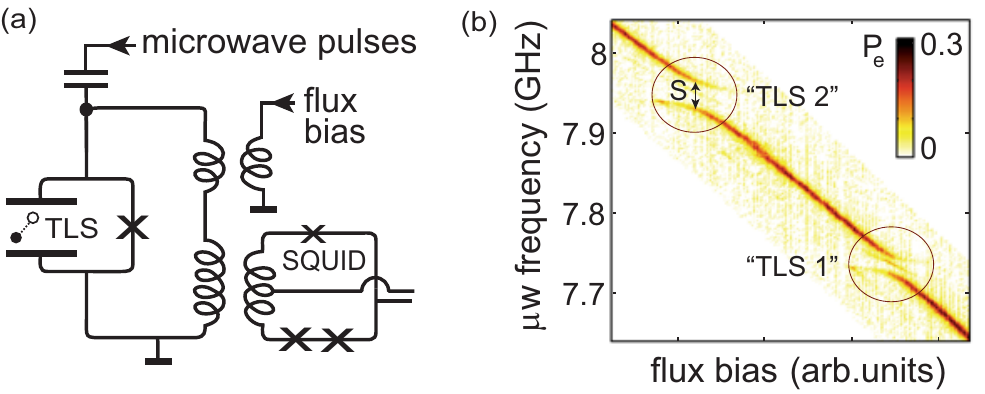} 
	\caption{(Color online) \textbf{(a)} Schematic of the phase qubit circuit. 
		\textbf{(b)} Qubit spectroscopy. Probability to measure the excited qubit state $P_e$ after a long (300 ns) microwave pulse of small amplitude 
		as a function of the qubit flux bias and microwave frequency. Two avoided level crossings, denoted "TLS 1" and "TLS 2", 
		are observed at $7.735$ GHz and $7.947$ GHz with magnitudes $S = 23$ MHz and $36$ MHz, respectively.} 
	\label{fig:schematic}
\end{figure}

Experiments on individual TLSs became possible with the advent of superconducting quantum bit (qubit) circuits~\cite{Martinis04a}. 
Individual TLSs can couple strongly to the qubit via their electric dipole moment when they are located in the dielectric of the thin ($\approx 2$ nm) tunnel barrier of the Josephson junctions forming the qubit. This coupling manifests itself as an avoided level crossing in the qubit spectrum at bias values for which a certain TLS energy splitting $\Delta E$ matches the energy difference between the two qubit states~\cite{Martinis05b}.
The coupling strength between a TLS and the qubit follows directly from the magnitude of the avoided level crossing $S$ as indicated in Fig. \ref{fig:schematic} (b).

Time-resolved experiments on phase qubits have demonstrated that an individual TLS can be manipulated using the qubit as a tool to both fully control and read out its state, and their possible use as a quantum memory has been demonstrated~\cite{martinis_quantummemory}. 
Recently~\cite{lisenfeld10}, we showed that there exists an effective qubit mediated coupling between TLSs and an externally applied electromagnetic ac-field.

\begin{figure*}[thb]
	\includegraphics[width=.9\textwidth]{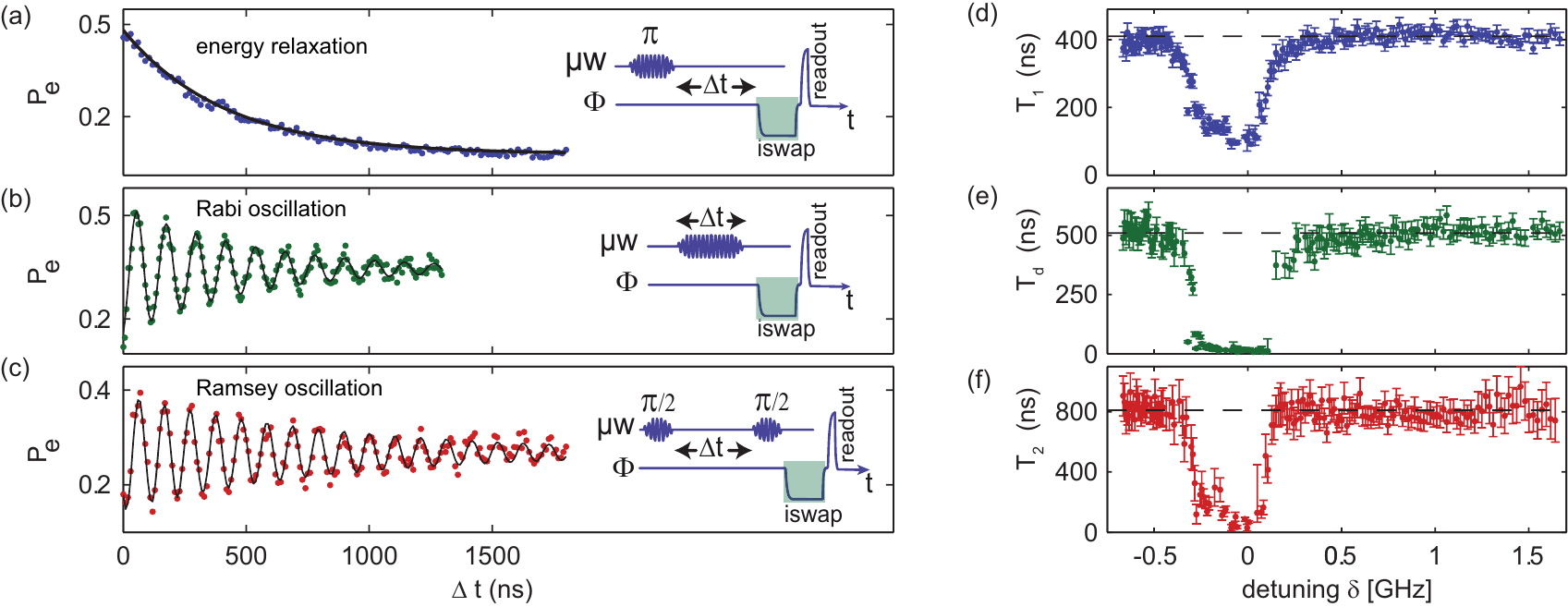}
	\caption{(Color online) Coherent response of TLS 1 to standard microwave pulse sequences as depicted in the insets. 
		\textbf{(a)} Relaxation of the excited TLS state after its preparation using a microwave $\pi$ pulse of duration 62 ns. Here, $\Delta t$ indicates the time delay between state preparation and readout. 
		\textbf{(b)} Rabi oscillation observed by driving the TLS resonantly for a duration $\Delta t$. 
		\textbf{(c)} Ramsey oscillation measured by varying the delay $\Delta t$ between two $\pi/2$ pulses (of duration 31 ns each) which were detuned from the TLS' resonance frequency by about 10 MHz. The data shown in panels \textbf{(a)}-\textbf{(c)} were obtained at a detuning $\delta = 502$ MHz.
		Panels \textbf{(d)}, \textbf{(e)} and \textbf{(f)} show the energy relaxation time $T_1^{TLS1}$, the Rabi decay time $T_d^{TLS1}$, and the dephasing time $T_2^{TLS1}$ 
		as a function of the detuning $\delta$ between TLS and qubit.} 
	\label{fig:biasdep}
\end{figure*}

In this work, we demonstrate that this coupling allows one to \emph{directly} control the quantum state of individual TLSs by coherent single-pulse resonant driving.
Since the qubit always remains detuned during TLS operation and merely acts as a detector to measure its resulting state, we can apply standard microwave pulse sequences at the TLS frequency in order to characterize its coherence properties. This supersedes the need for coherent qubit-TLS population exchange involving decoherence-limited excited qubit states, which would cause errors in the TLS manipulation, and enhances the possibility of using TLSs as the computational qubits~\cite{Nori06}. 

We use a phase qubit~\cite{Martinis04a} which consists of a superconducting LC - resonator, realized by shorting a capacitively shunted Josephson junction with an inductor as shown schematically in Fig. \ref{fig:schematic} (a). The Josephson junction acts as a nonlinear inductance providing anharmonicity to the oscillator, which is required to selectively excite transitions between the two lowest energy eigenstates. One can distinguish these qubit states by the rate at which one magnetic flux quantum enters the superconducting loop during application of a short (2 ns) readout flux pulse~\cite{martinisstatereadout}. The final flux state of the qubit is measured by an inductively coupled dc-SQUID. 

The sample used in this work had a qubit energy relaxation time $T_1^{q} \approx 110$ ns and a qubit dephasing time $T_2^{q} \approx 95$ ns and is otherwise identical to the sample presented in Ref.~\cite{Martinis07}. Microwaves are applied to the qubit via an on-chip planar transmission line being capacitively coupled to the shunting capacitor. 
During our experiments, the chip was maintained at a temperature of $35$ mK.
In several cool-downs of the sample studied, we observed between 3 and 4 TLSs coupling to the qubit with a strength larger then 10 MHz, which is about the spectroscopic resolution given by the linewidth of the resonance peak. Their energy splitting, coupling strength as well as coherence times changed after temperature cycles in which the superconducting transition temperature of aluminum $\approx$ 1.2 K was exceeded, but otherwise remained constant during several months of measurements \cite{Katz10}.

In order to use the qubit as a detector measuring the state of a certain TLS, it is first initialized in the ground state and biased at a flux at which both systems are sufficiently detuned such that their coupling can be neglected. A flux pulse then brings both systems into resonance, which gives rise to coherent oscillations within the coupled system. By choosing the flux pulse duration such that exactly half a period of oscillation occurs, the TLS state is mapped to the qubit state (under acquisition of a phase factor). 
Accordingly such a flux pulse is called an \emph{iswap} pulse. The duration of the iswap pulse, $T_\mathrm{swap} = S^{-1}/2$, corresponds to half the inverse magnitude of the avoided level crossing found spectroscopically. A subsequent readout of the qubit state then reflects the TLS population \cite{martinis_quantummemory}.

Direct manipulation of the TLS is straightforward: a microwave pulse applied to the qubit at the TLS resonance frequency $\nu_\mathrm{TLS} = \Delta E/h$ 
prepares its state via Rabi oscillation in a similar way as if the TLS would directly couple to the externally applied microwave. 
Similar to the situation we described in Ref.~\onlinecite{lisenfeld10}, this coupling to the driving field is mediated by a second order Raman process involving excited states of the qubit.

In Fig. \ref{fig:biasdep}, we present data obtained on TLS 1 whose resonance frequency was $\nu_\mathrm{TLS 1} \approx 7.735$ GHz and which was coupled to the qubit with a strength of $S_{1} \approx 23$ MHz. The energy relaxation time of the TLS is measured by applying a resonant microwave pulse of duration equal to half the Rabi period (a so-called $\pi$-pulse), which prepares the TLS in its excited state. By delaying the TLS readout sequence, which always consists of the iswap pulse followed by the qubit readout pulse, we observe exponential decay of the TLS excited state at a characteristic time $T_1^{TLS1}$ as shown in Fig. \ref{fig:biasdep} (a).
In Fig. \ref{fig:biasdep} (b), we show data obtained by applying a microwave pulse of varying duration to observe Rabi oscillations in the time domain, which decay at a characteristic time $T_d^{TLS1}$. Figure \ref{fig:biasdep} (c) shows Ramsey fringes created by applying two delayed $\pi/2$ pulses whose frequency was detuned from the exact TLS resonance by approximately 10 MHz. From the exponential decay of the Ramsey oscillations, we extract the dephasing time $T_2^{TLS1}$. 

To check whether the obtained coherence times are influenced by the coupling to the qubit, we repeated the measurements described above at different flux bias values. In Figs. \ref{fig:biasdep} (d)-(e), we plot the dependence of the coherence times on the detuning between qubit and TLS, $\delta = \nu_\mathrm{TLS} - \nu_\mathrm{qubit}$.
We see that the measured TLS coherence times do not depend on the detuning as long as $|\delta|$ is larger than a minimal value of about 500 MHz.
From these data we conclude that for large detuning, the TLS coherence is not limited by effects arising from the coupling to the qubit. 
For smaller detunings the qubit may be excited, leading to population of higher excited qubit states by multi-photon processes
due to the relatively large power of the applied microwave pulse. 
In addition, the time evolution will be more complicated as the coupling between the systems is no longer negligibly small~\cite{Mueller09}.

\begin{figure}[b]
	\includegraphics[width=.9\columnwidth]{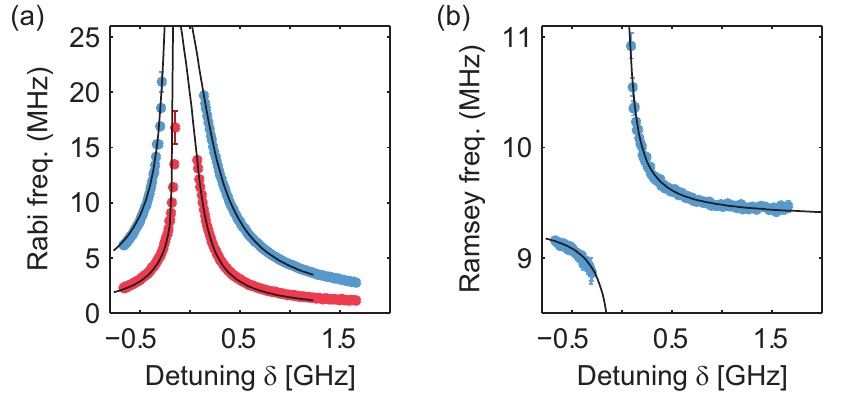} 
	\caption{(Color online) \textbf{(a)} Rabi frequency vs.\ detuning between TLS 1 and qubit. The two curves correspond to microwave powers differing by 9 dBm. 
		\textbf{(b)} Ramsey frequency vs.\ detuning between TLS 1 and qubit. Solid lines are calculated from theory.} 
	\label{fig:freqdep}
\end{figure}
Off-resonant, multi-photon excitation of the qubit to higher excited states is also the reason why the Rabi frequency of the TLS can not be increased arbitrarily by stronger driving. 
It is interesting to note that while this problem can in principle be circumvented by increasing the detuning, this counteracts faster TLS driving because for a given driving strength the TLS Rabi frequency decreases with increasing detuning as approximately $\nu_R \propto 1/\Delta$ as it is shown in Fig. \ref{fig:freqdep} (a). The theoretical curves in Fig. \ref{fig:freqdep} (a) indicate the Rabi frequencies of the system when driving resonantly with the TLS. The calculations included 5 higher levels in the qubit~\cite{lisenfeld10, bushev10}.

\begin{figure}
	\includegraphics[width =.8\columnwidth]{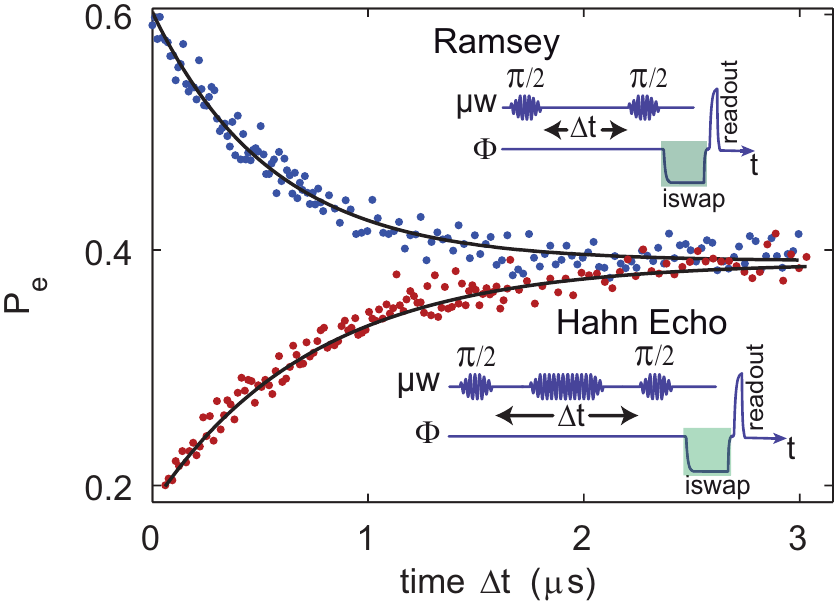} 
	\caption{(Color online) Signal amplitudes obtained from Ramsey (curve starting at $P_e$ = 0.6) and Hahn echo (curve starting at $P_e = 0.2$) sequences for TLS 2. 
		Solid lines are exponential fits resulting in a dephasing time of 551 $\pm$ 41 ns (Ramsey) and 743 $\pm$ 62 ns (Hahn echo), respectively.}
	\label{fig:spinecho}
\end{figure}
We note that the Ramsey frequency is not only determined by the drive detuning but is additionally modified by the coupling to the qubit (since the uncoupled states are no longer eigenstates), which gives rise to its dependence on the detuning $\delta$ as we show in Fig. \ref{fig:freqdep} (b).

We emphasize that for TLS 1, the average dephasing time $T_2^{TLS1} \approx 810 $ ns was very close to twice the average value of $T_1^{TLS1} \approx 410$ ns (indicated by dashed lines in Figs. \ref{fig:biasdep} (d) and (f). This is expected from a quantum system whose decoherence is limited purely by energy relaxation. Accordingly, we argue that this TLS was not coupled to any low-frequency noise sources which introduced dephasing on the observed time scale, such as fluctuating local fields affecting the TLS asymmetry $\Delta E$.

We measured a second TLS found in the sample during the same cool-down (TLS2). 
We obtained a similar dependence on the detuning parameter $\delta$ (data not shown here).
However, TLS 2 had a maximal dephasing time of $T_2^{TLS2} \approx 580$ ns whereas its energy relaxation time was $T_1^{TLS2} \approx 380$ ns. 
We found that application of a three-pulse echo sequence corrected for the excess dephasing in this TLS. We illustrate this in Fig. \ref{fig:spinecho}, where we compare the decay of the signal in a Ramsey protocol consisting of two resonant $\frac{\pi}{2}$ pulses to the Hahn echo sequence~\cite{Hahn} which adds a refocussing $\pi$ pulse in the middle.
The $T_2^*$ time obtained by the Hahn echo sequence is close to twice the energy relaxation time $T_1^{TLS2}$, from which we conclude that the dephasing mechanism acting on this particular TLS induces low-frequency energy fluctuations.

\begin{figure}
	\includegraphics[width=.95\columnwidth]{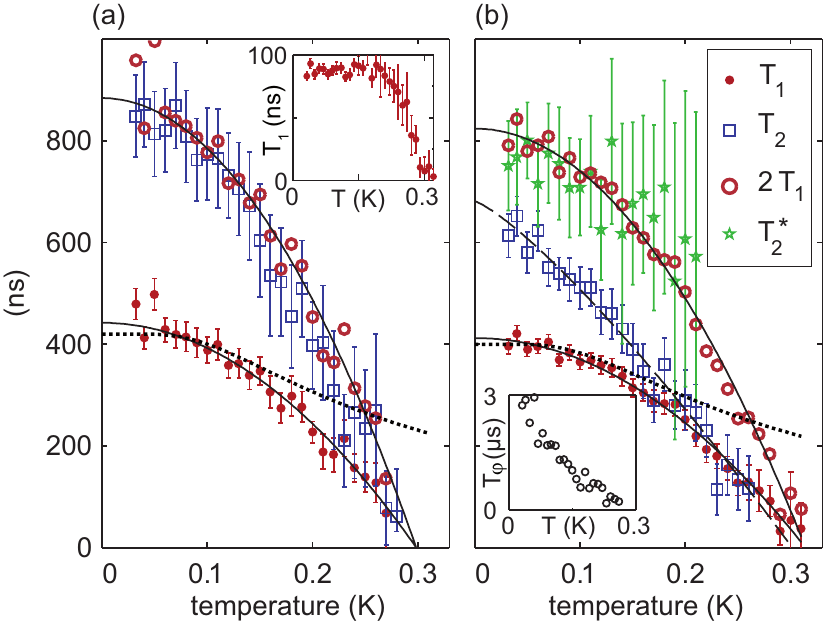} 
	\caption{(Color online) Temperature dependence of coherence times in \textbf{(a)} TLS 1 and \textbf{(b)} TLS 2. 
		The value of $2\cdot T_1$ is indicated by open circles. Solid lines are parabolas as described in the text, dotted lines are functions $T_1(T) \propto \tanh (\Delta E / 2 k_B T)$. 
		\textbf{Insets:} \textbf{(a)} Temperature dependence of the qubit $T_1$ time. \textbf{(b)} Extracted pure dephasing time $T_\varphi = 1/[T_2^{-1}- (2 T_2)^{-1}]$.
}
	\label{fig:tempdep}
\end{figure}

With the help of the established direct TLS manipulation procedure described so far, it is straightforward to measure the temperature dependence of TLS coherence times. As we showed in a previous work \cite{Lisenfeld07}, the coherence time of this phase qubit shows a weak temperature dependence up to a 
point where the thermal energy $k_B T$ approaches the qubit splitting $E_{|1\rangle}-E_{|0\rangle}$. To prevent the qubit from being thermally excited, we bias it above the studied TLS' resonance frequency at a chosen detuning of  $\delta = - 500$ MHz. 
In Fig. \ref{fig:tempdep}, we show the temperature dependencies of the TLSs' $T_1$ and $T_2$ times which were obtained by addressing the TLSs directly with the described pulse sequences.

We did not observe any indication that the TLSs' resonance frequency and coupling strength to the qubit varied with temperature.
For both TLSs, we observed an approximately quadratic decrease of the energy relaxation time $T_1$ with temperature. This is illustrated by the solid lines in Fig.~\ref{fig:tempdep} which are plots of the equation $T_1 (T) = T_1 (0) -  a T^{2}$, where a is a fitting parameter with the values $a = 4.96 \pm 0.71~ \mu s / K^2$ for TLS 1 and  $a = 4.18 \pm 0.36~ \mu s / K^2$ for TLS 2. 
Such behavior is not expected from a simple model, e.g., coupling the TLS to a bath of harmonic oscillators 
which would give $T_1(T) \propto \tanh (\Delta E / 2 k_B T)$~\cite{legget} as shown by the dotted lines in Fig~\ref{fig:tempdep}.
We stress that the $T_1^{q}$ time of the qubit shows a qualitatively different behavior as compared to the TLSs. This is illustrated by the data shown in the inset to Fig. \ref{fig:tempdep} (a), which was measured at a qubit splitting of $h\cdot 8.235$ GHz during the same cool-down.

We note that in TLS 1, the dephasing time $T_2$ remains close to twice its $T_1$ value in the whole temperature range. 
This TLS thus appears to remain immune to pure dephasing even at elevated sample temperatures.
In contrast, for TLS 2, which showed $T_2 < 2\cdot T_1$, the dephasing time is not well fitted by a parabolic temperature dependence. 
We illustrate this by the dashed line in Fig.~\ref{fig:tempdep} (b), which is a three-parameter fit resulting in the equation
$T_2 (T) = 681~ \mathrm{ns} -  3.01\ \mu\mathrm{s~K}^{-1.24}~ T^{1.24}$.
This can also be seen from the temperature dependence of the low-frequency noise contribution $T_{\varphi}$ as shown in the inset of Fig.~\ref{fig:tempdep} (b).
The Hahn-echo pulse suppressed excess dephasing in this TLS also at higher temperatures as shown in the same figure. 
However, the visibility of the echo signal decreased rapidly with temperature as the $T_1$ time became comparable to the duration of the microwave pulse sequence, 
in which we used a $\pi$-pulse of duration 62 ns.

In conclusion, we have demonstrated a new method to directly control TLSs in phase qubits, and applied it to measure TLS coherence times using standard pulse sequences.
Using this technique, we obtained the first data illustrating the temperature dependence of individual two-level systems in an amorphous solid. We observed an approximately quadratic decrease of the energy relaxation time $T_1$ with temperature in both measured TLSs, of which one did not show any excess dephasing. 

We are grateful to J. Martinis (UCSB) for providing the qubit used in this work.
We acknowledge financial support from the European projects SOLID and MIDAS and the US ARO under contract W911NF-09-1-0336.

\end{document}